\documentclass[
reprint, onlinecite, superscriptaddress,
amsmath,amssymb,
aps,prl
]{revtex4-2}

\usepackage{makecell}
\usepackage{graphicx}% Include figure files
\usepackage{dcolumn}% Align table columns on decimal point
\usepackage{bm}% bold math
\usepackage{longtable}
\usepackage{physics}
\usepackage{amsmath}
\usepackage{multirow}

\begin{document}
\title{Designing Spin-driven Multiferroics in Altermagnets}

\author{Ranquan Cao}
\affiliation{Key Lab of Advanced Optoelectronic Quantum Architecture and Measurement (MOE), School of Physics, Beijing Institute of Technology, Beijing 100081, China}%

\author{Ruizhi Dong}
\affiliation{Key Lab of Advanced Optoelectronic Quantum Architecture and Measurement (MOE), School of Physics, Beijing Institute of Technology, Beijing 100081, China}%

\author{Ruixiang Fei}
\email{rfei@bit.edu.cn}
\affiliation{Key Lab of Advanced Optoelectronic Quantum Architecture and Measurement (MOE), School of Physics, Beijing Institute of Technology, Beijing 100081, China}%

\author{Yugui Yao}
\email{ygyao@bit.edu.cn}
\affiliation{Key Lab of Advanced Optoelectronic Quantum Architecture and Measurement (MOE), School of Physics, Beijing Institute of Technology, Beijing 100081, China}%

\begin{abstract} 
%Spin-driven multiferroics often exhibit 
%large magnetoelectric coupling, with significant changes 
%in polarization under a magnetic field, typically found 
%in high-$Z$ magnetic insulators but with small polarization. 
Spin-driven multiferroics exhibit strong magnetoelectric coupling, 
with notable polarization changes under a magnetic field, but these
effects are usually limited to high-$Z$ magnetic insulators 
with low electronic polarization.
In this work, we introduce altermagnets as 
a promising platform for achieving strong magnetoelectric 
coupling in low-$Z$ systems with substantial polarization. 
This large polarization arises from a design principle 
that utilizes the Heisenberg-like exchange striction mechanism,
eliminating the reliance on spin-orbit coupling (SOC).
This approach enables the Kramers-degenerate antiferromagnetic 
phase derived from altermagnetic insulators to achieve substantial
polarization without spin splitting, providing a flexible
platform for regulating spin-splitting phenomena. 
Through first-principles simulations and an effective Landau-Ginzburg 
Hamiltonian, we demonstrate that materials in the LiMnO$_2$ family 
and strained RuF$_4$ family can achieve polarization
values exceeding 1.0 $\mu$$C/cm^2$, 
an order of magnitude larger than those found 
in SOC-driven multiferroics. Moreover, their magnetoelectric coupling
is one to two orders of magnitude 
stronger than that observed in conventional multiferroics 
and those driven by SOC. %, such as BiFeO$_3$ and TbMnO$_3$.
\end{abstract}

%\pacs{Valid PACS appear here}%PACS, the Physics Classification Scheme.
%\keywords{Suggested keywords} $Use showkeys class option if keyword display desired
\maketitle

\onecolumngrid

\textbf{\textit{Introduction}}-- 
Multiferroics, primarily involving the coexistence 
of magnetism and ferroelectricity, have garnered 
significant attention due to their intriguing physics 
and the potential to drive faster, smaller, 
energy-efficient data storage \cite{Eerenstein2006,Fiebig2016,Ramesh2007}. 
In particular, type II multiferroics, characterized by
a strong coupling between magnetic properties and 
magnetically induced ferroelectricity, are crucial for 
controlling electronic polarization using magnetic fields and vice versa.
\cite{Kimura2003,Hur2004}. 
To date, three primary approaches for generating type II multiferroic
have been identified, categorized into spin-orbit coupling (SOC)-related
and SOC-unrelated cases \cite{Fiebig2016,Tokura2014,Cheong2007,Dong2019}. 
The SOC-related phenomena involve mechanisms such as the
Dzyaloshinskii–Moriya (DM) interaction \cite{Katsura2005,Mostovoy2006} and 
spin-dependent metal–ligand $p$-$d$ hybridization \cite{Arima2007,Murakawa2010,Xiang2011}. 
In these cases, the polarization values 
are strongly correlated with the strength of the SOC,
which typically requires the presence of heavy elements. 
However, this relativistic effect is often weak, 
leading to relatively modest electronic polarization. 
For instance, the DM and inverse DM mechanisms 
induced ferroelectricity in orthorhombic TbMnO$_3$ \cite{Kimura2003} 
and CaMn$_7$O$_{12}$ \cite{Johnson2012}, generally 
yielding polarization values below 0.1 $\mu$$C/cm^2$, 
with a maximum reported polarization 
of 0.3 $\mu$$C/cm^2$ \cite{Johnson2012,Terada2016}.

The SOC-unrelated case refers to the Heisenberg-like
exchange striction mechanism, which requires a 
commensurate spin order and low symmetry in the 
specific chemical lattice, but does 
not involve SOC \cite{Sergienko2006B}. 
In contrast to the SOC-dependent mechanism, 
exchange striction in collinear magnets can 
be much more pronounced, leading to a 
substantial increase in polarization. 
For example, a polarization of around 
%approximately $P \approx
${0.1} \thickspace \mu C/ {cm}^{2}$ has been
reported in compressed orthorhombic TbMnO$_3$ \cite{Aoyama2014} 
when the DM mechanism transitions to an exchange striction mechanism. 
Unfortunately, only a limited number of oxide materials,
such as TbMn$_2$O$_5$ \cite{Hur2004} and 
orthorhombic HoMnO$_3$ \cite{Sergienko2006L,Picozzi2007,Lorenz2007},
have been identified to exhibit exchange striction-induced multiferroicity.
%have been identified as exhibiting multiferroicity induced by exchange striction.

Recent studies on altermagnetic antiferromagnet (AFM) \cite{Hayami2019,Smejkal2020Sci,Yuan2020} have unveiled
fascinating phenomena, including large spin-splitting 
\cite{Hayami2019,Yuan2020,Smejkal2022X,Ma2021,Mazin2021} 
and the $T$-odd spin Hall effect \cite{Gonzalez-Hernandez2021}. 
These phenomena are independent of SOC and are instead directly
driven by the collinear spin alignment that satisfies
the commensurate spin order required by the 
exchange striction mechanism \cite{Tokura2014}. 
Moreover, planar or bulk $d$-wave and $g$-wave altermagnetic materials,
which typically exhibit low symmetry (e.g., two-fold or three-fold) 
\cite{Smejkal2022X}, provide excellent platforms 
for exploring potential spin-induced multiferroicity 
via the exchange striction mechanism.
Additionally, the spin conservation and pronounced spin splitting
in altermagnets make it ideal for spintronic applications. 
%Additionally, the spin conservation and pronounced spin splitting characteristic of altermagnetism make it highly suitable for spintronic applications.
%The interplay between altermagnetic and multiferroic properties is highly desirable, as it enables efficient control of spin-related phenomena through electric or magnetic fields.
The combination of altermagnetic and multiferroic properties 
is highly beneficial, as it allows for efficient control 
of spin-related phenomena using electric or magnetic fields.

In this study, we propose that a spin-induced multiferroic %phase 
can be generated from the altermagnetic phase. 
%Our results show that, within the supercell structure, 
%the collinear multiferroic phase is not an altermagnetic phase but a Kramers-degenerate antiferromagnetic (AFM) phase, exhibiting spin-down and spin-up degeneracy, along with a significant spontaneous electronic polarization driven by the exchange striction mechanism. 
Our results show that, within the supercell structure, 
the collinear multiferroic phase is a Kramers-degenerate
AFM phase, exhibiting spin-down and spin-up degeneracy
and a significant spontaneous electronic polarization 
driven by exchange striction.
Due to the spin-induced ferroelectricity, electric 
or magnetic fields can rapidly switch between different
magnetic phases, such as ferromagnetic, altermagnetic, and 
Kramers-degenerate AFM, providing potential 
for various applications. 
%Using first-principles simulations and an 
%effective Landau-Ginzburg Hamiltonian, we demonstrate that materials in the LiMnO$_2$ and strained RuF$_4$ families can achieve polarization values exceeding 1.0 $\mu$$C/cm^2$.
Using first-principles simulations and an 
effective Landau-Ginzburg Hamiltonian, we reveal that LiMnO$_2$ and
strained RuF$_4$ materials can achieve polarization above 1.0 $\mu$$C/cm^2$
with magnetoelectric coupling far exceeding conventional 
and SOC-driven multiferroics.
%Additionally, the magnetoelectric coupling in these materials is one to two orders of magnitude stronger than in conventional multiferroics and SOC-related spin-driven systems.

\textbf{Spin-induced ferroelectricity and phase diagram}:
Motivated by the exploration of exchange striction-induced 
multiferroic properties in HoMnO$_3$ \cite{Sergienko2006L,Picozzi2007},
we designed a spin structure within a $2 \times 2$
supercell of an altermagnet. % illustrated in FIG. 1. 
In Fig. 1(a), the red and blue atoms represent spin-up 
and spin-down magnetic moments, respectively, 
connected by the $\left [ C_2 || C_{4z} \right]$ 
spin transformation. 
Here, the $C_2$ denotes a $180^{\circ}$ rotation 
in the spin-only space, while the $C_{4z}$ transformation
involves proper or improper fourfold rotations
in real space \cite{Smejkal2022X}. 
This spin transformation causes the spin-up and 
spin-down bands to split,
yet the electron polarization remains zero due to space 
inversion symmetry. We design the spin structure 
in a zigzag arrangement, with spins represented by 
the blue and red atoms in Fig. 1(b).
%We design the spin structure in 
%a zigzag pattern, with spins arranged as shown 
%by the blue and red atoms in FIG. 1(b). 
%Notably, the spin-up and spin-down sublattices are related by the spin transformation
Notably, the spin-up and spin-down sublattices 
are connected by the spin transformation $\left [ C_2 || t \right]$, 
where $t$ represents a half-unit translation symmetry in real space. 
%This symmetry ensures that spin-up and spin-down bands remain degenerate throughout the entire reciprocal space, categorizing the structure as a Kramers-degenerate AFM
This symmetry maintains the degeneracy of spin-up and spin-down
bands across the entire reciprocal space, 
classifying the structure as a 
Kramers-degenerate AFM (type II AFM).

\begin{figure}[!]
\includegraphics[width=10cm]{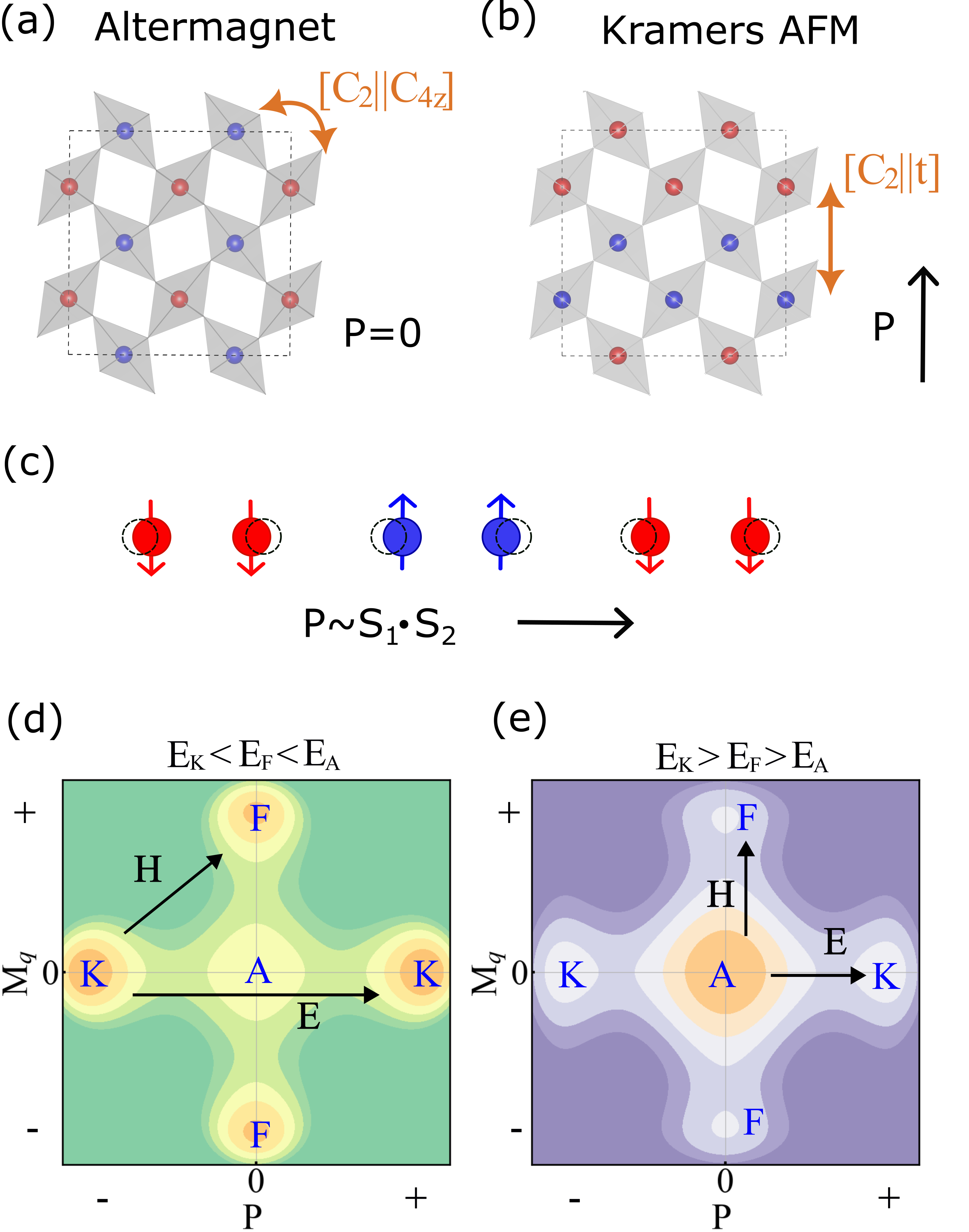}
\caption{ 
Atomic structure showing only the magnetic atoms, 
with the spin pattern in (a) representing the 
altermagnetic phase and in (b) representing 
the Kramers-degenerate AFM phase. (c) Spin pattern 
along the y-direction in (b) showing a Peierls-like 
phase transition. Phase diagrams for the
Kramers-degenerate AFM (d) and altermagnetic (e) 
as ground states, with K, A, and F representing Kramers-degenerate AFM, altermagnetic, and ferromagnetic phases, respectively.
} \label{Figure1}
\end{figure}

We observe that the electronic polarization of 
this spin structure is oriented along the y-axis,
as shown in Fig. 1(b). 
This polarization arises from two contributions:
one due to the inversion symmetry breaking induced
by the spin pattern, and the other from the ionic component.
As depicted in Fig. 1(c), the spin pattern disrupts
space inversion symmetry. Concurrently, the crystal-field-modified 
super-exchange interaction facilitates an attractive 
interaction between parallel spins and a repulsive 
interaction between antiparallel spins. 
This mechanism drives the ions to undergo a Peierls-like 
structural phase transition, leading to a moderate 
polarization along the y-axis. If the free energy of 
this Kramers-degenerate AFM configuration is 
significantly lower than that of the altermagnetic phase,
the material could exhibit stable multiferroic
behavior with strong magnetoelectric coupling.

Regardless of which phase is the ground state,
the Hamiltonian can be expressed in the Landau-Ginzburg form:
\begin{align}
\begin{split}
H=\frac{1}{2}{a}_{p}(\eta ){P}^{2}+\frac{1}{4}{b}_{p}(\eta ){P}^{4}+\frac{1}{6}{c}_{p}(\eta ){P}^{6} +\frac{1}{2}{a}_{q}(\eta ){M}_{q}^{2} \\+\frac{1}{4}{b}_{q}(\eta ){M}_{q}^{4}+\frac{1}{6}{c}_{q}(\eta ){M}_{q}^{6} 
   +\frac{1}{2}{\lambda }(\eta ){M}_{q}^{2}{P}^{2}-EP
\end{split}
\end{align}
Here, we focus on the spin-induced polarization $P$. 
For each strain $\eta$ and external electric field $E$,
the ground state is determined by minimizing the free 
energy with respect to $P$ and $M_q$.
The magnetic order parameter $M_q$ is defined as: 
${M}_{q}=\frac{1}{N}\textstyle\sum_{j}{e}^{iq\cdotp {R}_{j} }\left \langle {S}_{j}\right \rangle$, 
where $\left \langle {S}_{j}\right \rangle$ is 
the thermodynamic average of the normalized spin 
at the site $R_j$ and $N$ is the number of spins.
Taking MnF$_2$ as an example, $q=\pi (0,0,\frac{2}{c})$
stands for the altermagnetic phase, and 
$q=\pi (\frac{2}{a},\frac{1}{a},\frac{1}{c})$ 
corresponds to the Kramers-degenerate AFM. 
The parameter $\lambda (\eta )$ denotes the 
magnetoelectric coupling strength under the specified strain $\eta$.
Using this Hamiltonian, a phase diagram can be constructed 
to depict the transitions between the Kramers-degenerate
AFM phase, the altermagnetic phase, and the ferromagnetic phase.

Two primary scenarios can be identified: 
%one in which the Kramers-degenerate AFM phase
%represents the ground state, and another in which 
%the altermagnetic phase constitutes the ground state, 
%as shown in Fig. 1(d) and 1(e), respectively.
one where the Kramers-degenerate AFM phase is 
the ground state, and another where the altermagnetic 
phase serves as the ground state. 
%, as shown in Fig. 1(d) and 1(e), respectively.
In Fig. 1(d), the spontaneously polarized Kramers-degenerate 
AFM serves as the ground state, allowing the electric field 
to switch the negatively polarized state to a positively 
polarized structure. Concurrently, an applied magnetic field
can drive a transition from the Kramers-degenerate AFM 
to a non-polarized ferromagnetic phase, 
leading to strong magnetoelectric coupling.
%In the second scenario, illustrated in Fig. 1(e), the altermagnetic phase constitutes the ground state. Upon the application of an electric field, the system undergoes a transition from the altermagnetic phase to a Kramers-degenerate AFM phase. 
In the second scenario, shown in Fig. 1(e), the altermagnetic 
phase is the ground state. Upon the application of an electric field,
the system transitions from the altermagnetic phase to 
a Kramers-degenerate AFM phase.
%While the net magnetization remains invariant
While the net magnetization remains unchanged, 
the suppression of spin splitting results in the 
vanishing of phenomena typically associated with 
altermagnets, such as the $T$-even symmetric spin Hall effect.
%Similarly, the application of a magnetic field can reduce spin splitting within the altermagnetic phase, resulting in the transformation of the spin Hall effect into the anomalous Hall effect.
Similarly, applying a magnetic field can diminish 
spin splitting within the altermagnetic phase, 
causing the spin Hall effect to transform into 
the anomalous Hall effect.

\textbf{\textit{Materials realizations}}: 
Non-collinear spin alignment in antiferromagnets can 
spontaneously break inversion symmetry, leading to 
spin-driven ferroelectricity \cite{Lim2018}. However, 
the DM interaction, arising from SOC, can also promote 
non-collinear spin configurations, making it challenging to
disentangle the contributions of non-SOC mechanisms to 
spin-driven ferroelectricity \cite{Katsura2005,Mostovoy2006}. 
Moreover, these systems generally display weak electronic 
polarization, restricting their utility in 
multiferroic applications. 
To address this, we focus on collinear systems to 
investigate spin-driven ferroelectricity,
magnetoelectric coupling, and potential phase transitions. 
Our first-principles calculations \cite{SI} reveal 
that numerous materials identified as $g$-wave or $d$-wave 
altermagnets, which possess lower crystal symmetry
compared to $i$-wave altermagnets, can transition into 
Kramers-degenerate antiferromagnets, functioning 
as spin-induced multiferroics.

A representative example is the LiMnO$_2$ family, 
which includes NaMnO$_2$ compounds. These structures 
can be described as Li (or Na) ions intercalated 
into MnO$_2$ frameworks, a hallmark of ionic 
battery materials \cite{Uyama2018}. 
%First-principles calculations indicate that 
%several experimentally observed phases of LiMnO$_2$ exhibit 
%altermagnetic behavior. Notably, the orthorhombic LiMnO$_2$ (Pmmn) and NaMnO$_2$ (I4$_1$/amd) phases serve as examples. 
First-principles calculations reveal that 
several experimentally observed phases 
of LiMnO$_2$ family display altermagnetic behavior,
with the orthorhombic LiMnO$_2$ (Pmmn) 
and NaMnO$_2$ (I4$_1$/amd) phases being notable examples.
However, achieving a transition from the altermagnetic phase to a Kramers-degenerate phase in these materials necessitates lattice compression or strain exceeding $8\%$, posing substantial experimental challenges.

\begin{figure}[!]
\includegraphics[width=10cm]{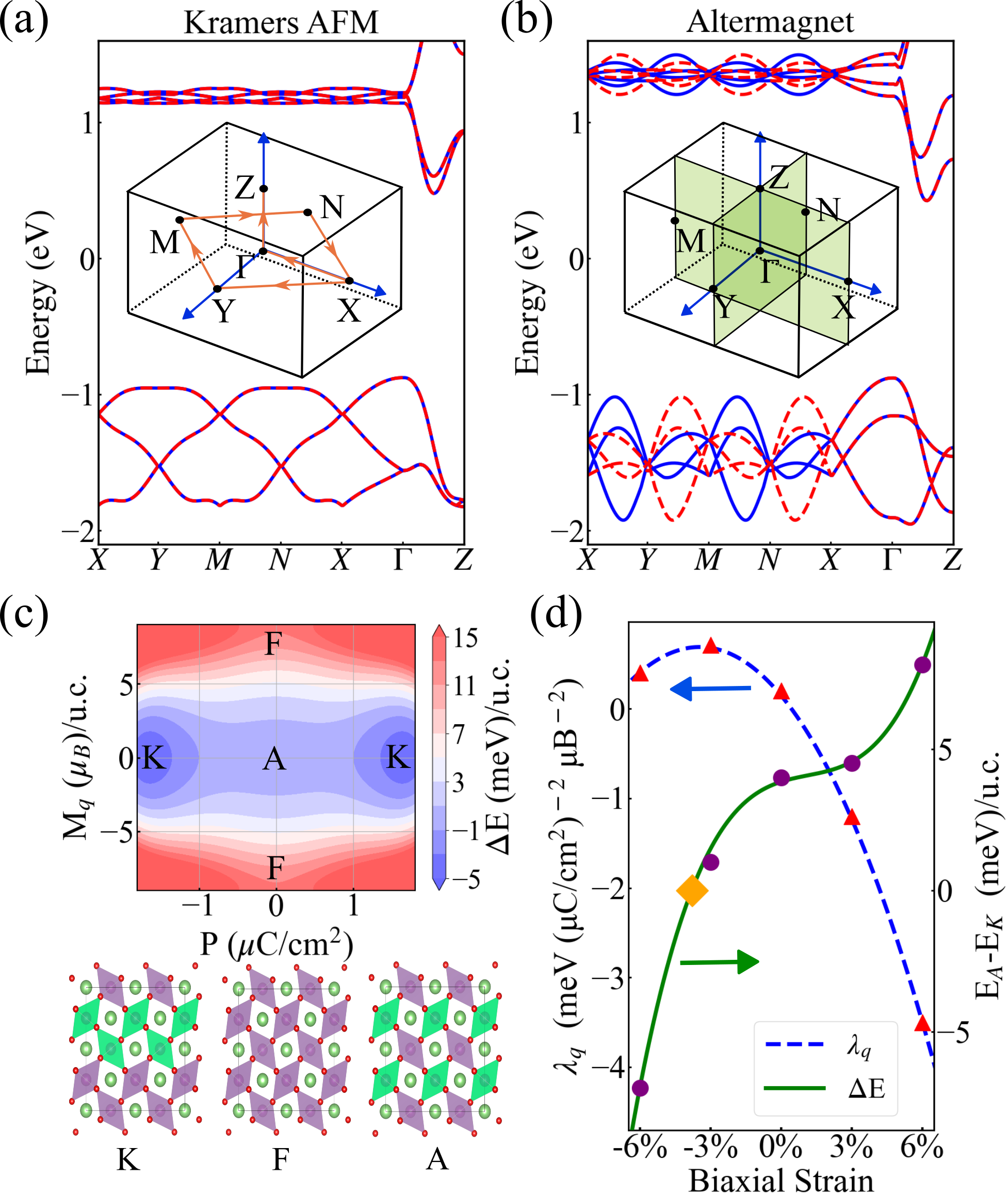}
\caption{
(a) Band structure of the Kramers-degenerate AFM phase and (b) the altermagnetic phase. The green-highlighted plane in (b) indicates the spin-degenerate plane of the altermagnet. (c) Phase diagram of LiMnO$_2$, showing the Kramers-degenerate AFM (K), altermagnetic (A), and ferromagnetic (F) phases. (d) Magnetoelectric coupling constant (blue dashed line) and the energy difference between the A and K phases (green solid line), with the transition point indicated by the diamond symbol.}
\label{Fig2}
\end{figure}

The DFT+U calculations \cite{SI,Kresse1996,Dudarev1998,perdew1996,Kresse1999}
identify the Kramers-degenerate phase in the stable
P2$_1$2$_1$2 configuration as the ground state.
Fig. 2(a) presents the corresponding Brillouin zone 
and the band structure of the Kramers-degenerate AFM. 
%The degeneracy of the spin-up and spin-down bands 
%arises from the half-unit translation operator 
%linking the spin-down and spin-up sublattices,
%match to the mechanism shown in FIG. 1(b). 
The spin-up and spin-down band degeneracy originates 
from the half-unit translation operator connecting 
the spin-down and spin-up sublattices, consistent 
with the mechanism illustrated in Fig. 1(b).
In contrast, in the altermagnetic phase, 
the operator $\left [ C_2 || C_{2z} \right]$ connects 
the spin-down and spin-up sublattices, resulting in spin 
splitting outside the high-symmetry plane, 
as indicated by the green shading in the inset of Fig. 2(b). 
This behavior classifies the system as a planar $g$-wave 
altermagnet.  

%We obtained the phase diagram from the effective Hamiltonian,
% for LiMnO$_2$, 
%as given in Eq. (1), by calculating various spin configurations using the DFT+U method \cite{SI}.
We derived the phase diagram from the effective Hamiltonian
in Eq. (1) by calculating various spin configurations 
using the DFT+U method \cite{SI}.
The K, A, and F phases represent the Kramers-degenerate AFM, 
altermagnetic, and ferromagnetic phases, respectively, 
with green and purple octahedra in the lower panel 
of Fig. 2(c) denoting spin-up and spin-down sites.
Our calculations reveal that the ground state of 
this structure is a Kramers-degenerate AFM with 
a polarization of $P = {0.1} \thickspace \mu C/ {cm}^{2}$,
arising from both the spin pattern and a spin-induced 
Peierls-like structural transition. 
When calculating the polarization for the
Kramers-degenerate spin configuration with 
fixed atomic positions, excluding the Peierls-like transition,
we find a value of 0.8 $\mu C/ {cm}^{2}$, nearly an
order of magnitude larger than that from SOC-related mechanisms.

%The electromagnetic coupling constant, $\lambda $, can be derived from Eq. (1), which quantifies the overall ability to control electronic polarization with an magnetic field, or vice versa. 
The electromagnetic coupling constant, $\lambda$, 
is derived from Eq. (1) and quantifies the overall 
ability to control electronic polarization with a 
magnetic field, or vice versa.
%The blue dashed line in 
Fig. 2(d) illustrates how the electromagnetic coupling 
constant varies with biaxial strain or compression.
%the variation of the electromagnetic coupling constant with biaxial strain or compression. 
%It is intuitive that this constant would be influenced by strain, as strain alters atomic distances and the Mn-O-Mn angle, both of which can significantly affect the Heisenberg exchange interaction. 
Intuitively, the strain would affect this constant, as it alters atomic distances and the Mn-O-Mn angle, both of which can significantly influence the Heisenberg exchange interaction.
The intrinsic magnetoelectric coupling near the
Kramers-degenerate antiferromagnetic phase is approximately
0.1 $meV\cdot {\left ( \frac{\mu C}{{cm}^{2}}\right )}^{-2}{\mu }_{B}^{-2}$,  
which is one to two orders of magnitude greater than
that driven by conventional or SOC mechanisms \cite{Edstrom2020L,Gupta2022,Shah2012}. 
We find that the magnitude of the magnetoelectric
coupling can increase by an order of magnitude 
under moderate strain, such as a 3$\%$ lattice strain. 
%We find that the absolute value of the magnetoelectric coupling can increase by an order of magnitude under moderate strain, such as a 3$\%$ lattice strain. 
Notably, a compressive strain of 4$\%$ or greater 
can shift the ground state from a Kramers-degenerate 
multiferroic to an altermagnetic phase,
as marked by the yellow diamond symbol on the
green line in Fig. 2(d).

Another notable example is the XF$_4$ 
family (X=Ru, V, Os), crystallizing in
the $P2_1/c$ space group and has been recognized
as experimentally stable van der Waals-bonded
bulk crystals \cite{Casteel1992,Becker1990}. 
This structure has the spin group $^22/^2m$,
%which indicate that the spin-up and spin-down sublattice are connected by the rotation transformation
where spin-up and spin-down sublattices 
are related by the transformations
$\left [ C_2 || C_{2z} \right]$ and 
$\left [ C_2 || M_{z} \right]$. 
%Since the $C_{2z}$ axis and $M_z$ plane are orthogonal to each other, and no mirror plane exists parallel to the $C_{2z}$ axis, there is only one spin-splitting off-nodal plane: the $\Gamma MK$ plane, as shown in Fig. 3(a). 
With the $C_{2z}$ axis and $M_z$ plane being 
orthogonal and no mirror plane parallel to the 
$C_{2z}$ axis, the only spin-splitting off-nodal 
plane is the $\Gamma MK$ plane, 
as shown in Fig. 3(a).
%This characteristic sets it apart from the typical $d$-wave altermagnet. Notably, the band structure along the $\Gamma Z$ line remains spin-degenerate (FIG. 3(b)), a result of the symmetry enforced by the $C_{2z}$ rotation axis.  
This feature distinguishes it from typical 
$d$-wave altermagnets. Specifically, the band
structure along the $\Gamma Z$ line remains 
spin-degenerate (Fig. 3(b)) due to the symmetry 
enforced by the $C_{2z}$ rotation axis.

\begin{figure}[!]
\includegraphics[width=10cm]{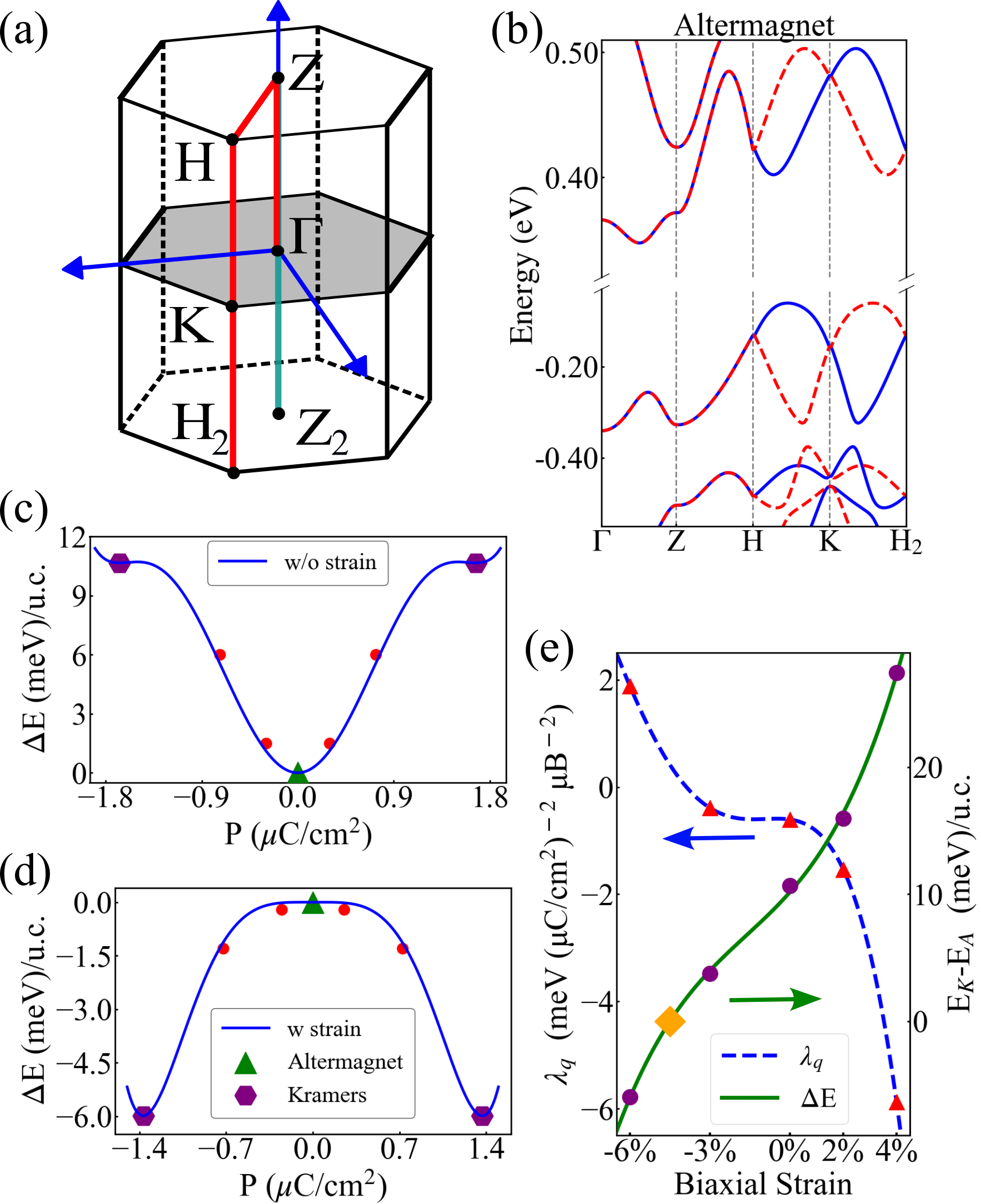}
\caption{
(a) Brillouin zone of the XF$_4$ family, with the spin-degenerate plane highlighted. (b) Band structure of the altermagnetic phase. (c) Energy potential for the $M=0$ state in the effective Hamiltonian for the intrinsic case. (d) Energy potential under 6$\%$ biaxial compression. (e) Magnetoelectric coupling constant (blue dashed line) and energy difference between the altermagnetic and Kramers-degenerate phases (green solid line) as a function of strain.}
\label{Fig3}
\end{figure}

Similar to the LiMnO$_2$ family, 
the Kramers-degenerate AFM phase in these 
materials is noncentrosymmetric, exhibiting 
a spontaneous polarization of up to 
1.8 $\mu C/{cm}^2$, as represented by the
hexagonal point in Fig. 3(c).
%This polarization surpasses that of hallmark multiferroics such as TbMnO$_3$ and TbMn$_2$O$_5$ by one to two orders of magnitude 
This polarization surpasses that of benchmark 
spin-induced multiferroics such as TbMnO$_3$ 
and TbMn$_2$O$_5$ by one to two orders of 
magnitude \cite{Kimura2003,Hur2004}. 
%We investigate the energy transition between the Kramers-degenerate and altermagnetic phases by rotating the spin alignment, resulting in distinct polarization states. 
We examine the energy transition between the 
Kramers-degenerate and altermagnetic phases by 
rotating the spin alignment, leading to distinct 
electronic polarization states.
The ground state is found to correspond to 
the altermagnetic phase, which is 10 $meV$ 
lower in energy per primitive cell than 
the Kramers-degenerate phase.

With in-plane lattice compression of the 
layered structure, the Kramers-degenerate 
AFM phase can become the ground state. 
As shown in Fig. 3(d), the double-well 
potential under 6$\%$ biaxial compression 
indicates the Kramers-degenerate phase as 
the ground state, featuring large 
spontaneous polarization.
%the  Kramers-degenerate phase  as its ground state with large spontaneous polarization. 
%We observe that biaxial compression greater than 4$\%$ triggers a phase transition from the altermagnetic phase to the Kramers-degenerate phase, as indicated by the green line in FIG. 3(e). 
Biaxial compression over 4$\%$ induces 
a phase transition from the altermagnetic 
to the Kramers-degenerate phase, as shown
by the green line in Fig. 3(e).
By deriving the effective Hamiltonian in Eq. (1),
we calculate the magnetoelectric coupling 
constant under various strain conditions. 
%The magnetoelectric coupling constant of the Kramers-degenerate AFM phase can reach up to approximately 2 $meV\cdot {\left ( \frac{\mu C}{{cm}^{2}}\right )}^{-2}{\mu }_{B}^{-2}$ under moderate strain or compression, representing an enhancement by an order of magnitude compared to the intrinsic value. 
Under moderate strain or compression, the 
magnetoelectric coupling constant of the 
Kramers-degenerate AFM phase can reach around 
2 $meV\cdot {\left ( \frac{\mu C}{{cm}^{2}}\right )}^{-2}{\mu }_{B}^{-2}$, 
an order of magnitude higher than its 
intrinsic value.
%Furthermore, we observe that the two-dimensional form of the XF$_4$ material family exhibits similar properties, including a large magnetoelectric coupling constant 
Additionally, the two-dimensional (2D) form
of the XF$_4$ family displays similar 
properties, including a high 
magnetoelectric coupling constant \cite{SI}.

\begin{figure}
\includegraphics[width=10cm]{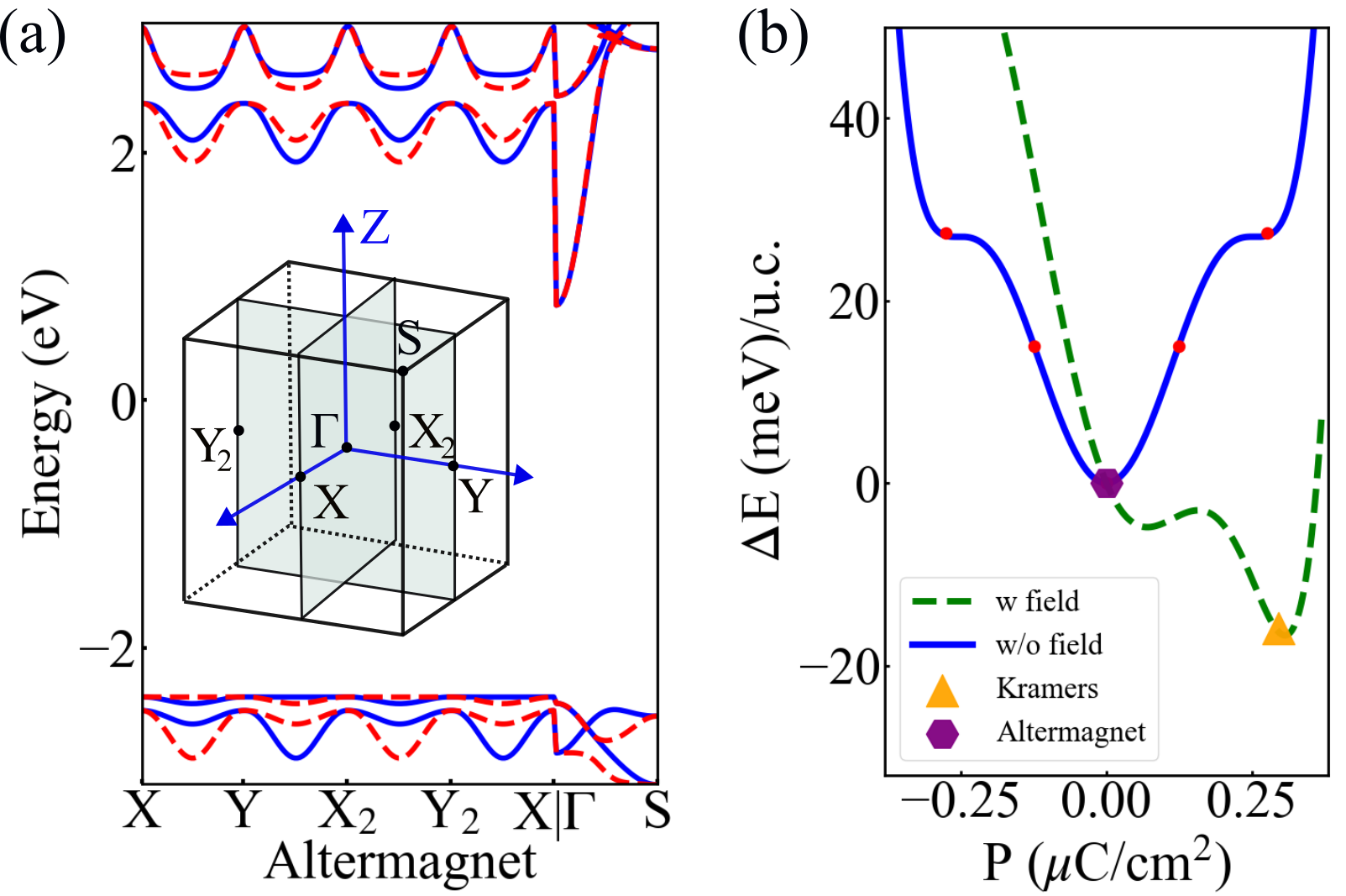} %width=8.6cm
%scale=0.22
\caption{(a) Band structure of the altermagnetic phase of MnF$_2$, with the Brillouin zone highlighting the spin-degenerate plane. (b) Energy potential of the zero total magnetic moment state for the intrinsic case (solid line) and under an applied in-plane electric field (dashed line).  
} 
\label{Fig4}
\end{figure}
%\twocolumngrid\

%Finally, we demonstrate that, in certain cases, the Kramers-degenerate phase is significantly higher in energy than the ground-state altermagnetic phase, which impedes the transition from the altermagnetic phase to the Kramers-degenerate phase through strain or compression. 
Finally, we show that in specific cases, 
the Kramers-degenerate phase is substantially 
higher in energy than the ground-state 
altermagnetic phase, thereby hindering 
strain- or compression-driven transitions 
between the two phases.
%This behavior is observed in several antiferromagnetic materials, including MnO$_2$ and members of the MnF$_2$ fluoride family. For instance, MnF$_2$ has been experimentally confirmed to adopt the rutile structure with the $P4_2/mnm$ space group \cite{Stout1942,Baur1971}. 
This behavior is evident in several 
antiferromagnetic materials, such as MnO$_2$ 
and members of the MnF$_2$ fluoride family. Notably,
MnF$_2$ has been experimentally verified to 
crystallize in the rutile structure with the
$P4_2/mnm$ space group \cite{Stout1942,Baur1971}. 

%The first-principles calculations show that the altermagnetic phase is approximately 30 $meV$ lower in energy per primitive cell compared to the Kramers-degenerate phase. Its altermagnetic phase is classified under the spin group $^24/^1m^2m^1m$, indicative of a planar $d$-wave type. 
First-principles calculations reveal that 
the altermagnetic phase is about 30 $meV$
lower in energy per primitive cell than the 
Kramers-degenerate phase. This altermagnetic 
phase belongs to the spin group $^24/^1m^2m^1m$,
characteristic of a planar $d$-wave type.
As illustrated in Fig. 4(a), the combined rotation
operation $\left [ C_2 || C_{4z} \right]$ and 
mirror operation $\left [ C_2 || M_{z} \right]$ 
guarantee the existence of two spin-degenerate
planes within the Brillouin zone.
%ensure the presence of two spin-degenerate planes within the Brillouin zone.

Fig. 4(b) shows the energy potential between 
the Kramers-degenerate and altermagnetic phases.
In the intrinsic case, the ground state 
corresponds to the altermagnetic phase.
First-principles calculations reveal that 
the polarization of the Kramers-degenerate 
phase is 0.25 $\mu C/{cm}^2$. 
Furthermore, we find that it is not possible
to stabilize the Kramers-degenerate phase 
as the ground state under moderate strain or
compression, such as a 15$\%$ lattice 
expansion or compression \cite{SI}. 
Instead, the phase transition between the 
Kramers-degenerate and altermagnetic phases 
can be regulated by an external electric field,
as outlined in Eq. (1), due to the distinct 
electronic polarization of these phases.
%controlled by an external electric field, as described by Eq. (1), given that these phases are distinguished by their electronic polarization.

\textbf{\textit{Giant magnetoelectric coupling}}:
In the proposed mechanism, ferroelectricity
originates from the spin alignment induced by
exchange striction. This spin-dependent 
ferroelectric polarization exhibits strong 
sensitivity to external magnetic fields, 
enabling significant modulation and yielding 
a large magnetoelectric coupling constant.
%allowing for substantial modulation and resulting in a giant magnetoelectric coupling constant. 
Our calculations indicate that both LiMnO$_2$ 
and strained RuF$_4$ family exhibit significant 
magnetoelectric coupling.%, as derived from Eq. (1). 
%As shown in Fig. 5, when converting the units to $mV/(Oe\cdot cm)$ for experimental comparison \cite{Eerenstein2006}, we find that these materials display a magnetoelectric coupling constant that is at least an order of magnitude larger than those observed in conventional multiferroics and SOC-induced multiferroics.
As shown in Fig. 5, converting the units to $mV/(Oe\cdot cm)$ for experimental comparison \cite{Eerenstein2006} reveals that these materials exhibit a magnetoelectric coupling constant at least an order of magnitude higher than those found in conventional and SOC-induced multiferroics.

\begin{figure}
\includegraphics[width=10cm]{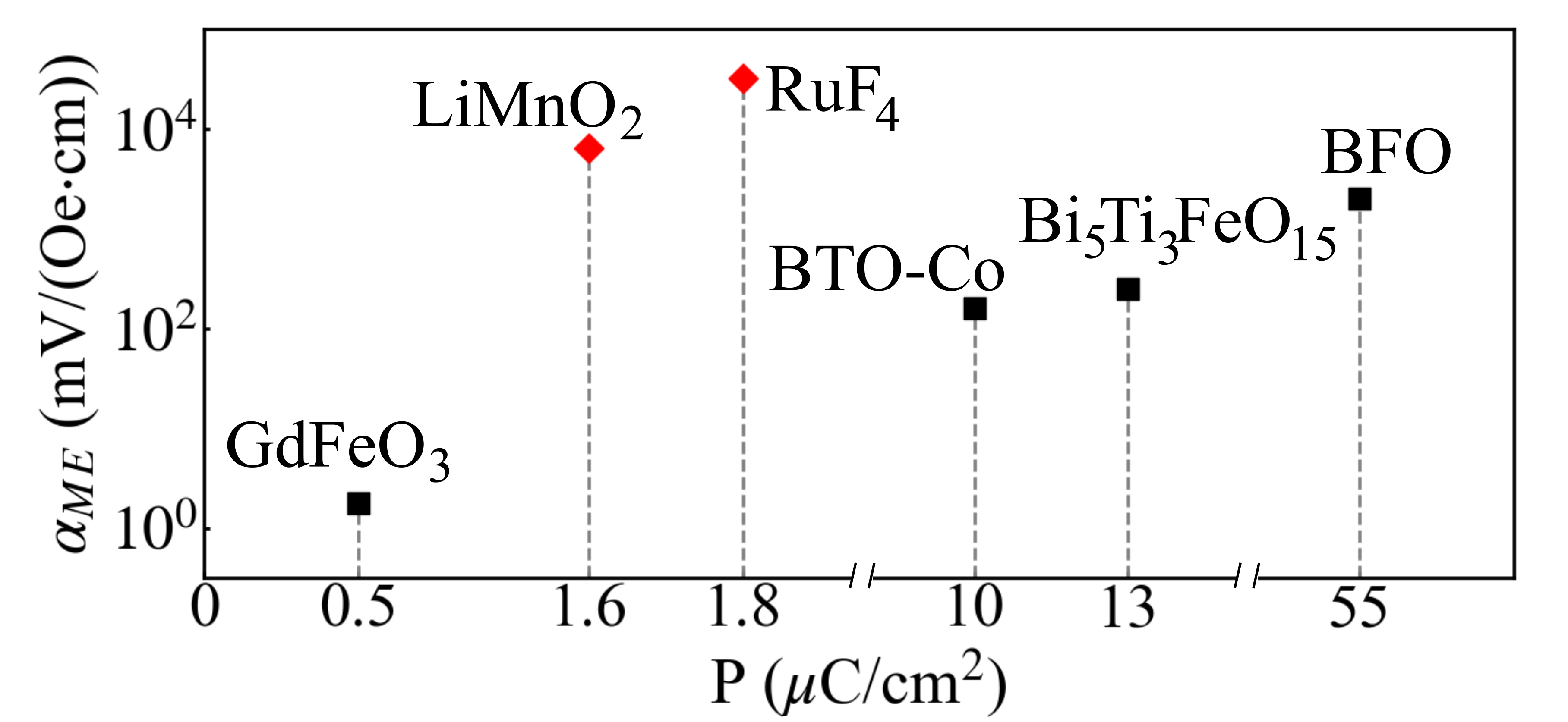} %width=8.6cm
%scale=0.22
\caption{Comparison of the magnetoelectric coupling constants for our calculated LiMnO$_2$ $(P2_12_12)$ and strained RuF$_4$, alongside representative multiferroics: BFO \cite{Gupta2022}, BTO-Co \cite{Park2008}, Bi$_5$Ti$_3$FeO$_{15}$ \cite{Zhao2014}, and GdFeO$_3$  \cite{Shah2012,Zhao2017}.  
} 
\label{Fig5}
\end{figure}

For comparison, the ferroelectricity induced
by the DM interaction in TbMnO$_3$ serves as 
a prototypical example of a spin-orbit coupling 
(SOC)-related mechanism  \cite{Kimura2003}. 
Although experimental values are unavailable, 
its magnetoelectric coupling is known to
be weaker than that of BiFeO$_3$ (BFO).
%While experimental values are not available, its magnetoelectric coupling is smaller than that of BiFeO$_3$ (BFO). 
Similarly, GeFeO$_3$, another material 
exhibiting SOC-driven ferroelectricity  
\cite{Shah2012,Zhao2017}, exhibits a 
magnetoelectric coupling constant two to three 
orders of magnitude lower than those of the 
materials we propose.
% has a magnetoelectric coupling constant that is two to three orders of magnitude smaller than those of the materials we propose. 
While the polarization in BFO primarily arises
from atomic displacement, its large
magnetoelectric coupling is generally 
attributed to the DM interaction \cite{Ederer2005}. 
%In contrast, our proposed system, which operates via the exchange striction mechanism, exhibits a magnetoelectric coupling constant that is at least an order of magnitude larger than that of BFO.
In contrast, our proposed system, based on 
the exchange striction mechanism, exhibits a
magnetoelectric coupling constant an order 
of magnitude higher than that of BFO.

In summary, we propose that the recently explored 
altermagnetic systems offer an ideal platform for
designing spin-driven multiferroics through 
the exchange striction mechanism.
%In summary, we propose that the recently studied altermagnetic systems can serve as an ideal platform for designing spin-driven multiferroics via the exchange striction mechanism. 
This principle is universal for altermagnetic systems, 
as they inherently meet the criteria for 
%satisfy the requirements of
the exchange striction mechanism:
a commensurate spin order and low symmetry 
within the specific chemical lattice. 
We find that the collinear multiferroic phase is 
a Kramers-degenerate AFM, characterized by spin-up 
and spin-down degeneracy, along with a significant 
spontaneous polarization driven by exchange striction. 
Through first-principles simulations and an effective
Landau-Ginzburg Hamiltonian, we show that materials 
such as the LiMnO$_2$ and strained RuF$_4$ families, 
classified as low-$Z$ multiferroics, can achieve
polarization values exceeding 1.0 $\mu C/{cm}^2$. 
These materials display magnetoelectric coupling 
constants one to two orders of magnitude stronger 
than those in conventional multiferroics and 
SOC-driven systems. Moreover, moderate strain 
can effectively switch between the altermagnetic 
and Kramers-degenerate multiferroic phases, 
offering further opportunities to tune 
the magnetoelectric coupling.

\begin{acknowledgments}
This work is supported by the National Natural Science Foundation of China Grant No. 12204035.
\end{acknowledgments}

\bibliography{main}
\end{document}